\documentclass[conference]{IEEEtran}
\IEEEoverridecommandlockouts

\usepackage{cite}
\usepackage{amsmath,amssymb,amsfonts}
\usepackage{algorithmic}
\usepackage{graphicx}
\usepackage{textcomp}
\usepackage{xcolor}
\usepackage{dirtytalk}

\def\BibTeX{{\rm B\kern-.05em{\sc i\kern-.025em b}\kern-.08em
    T\kern-.1667em\lower.7ex\hbox{E}\kern-.125emX}}

\begin{document}

\title{AI as a Tool for Fair Journalism: Case Studies from Malta\\
\thanks{This work is part of the project 'Exploring Visual Bias in News Content using Explainable AI' (NBxAI) and is financed by the Malta Council for Science \& Technology, for and on behalf of the Foundation for Science and Technology, through the FUSION: R\&I Research Excellence Programme.
This paper will be published in the IEEE 2024 Conference on Artificial Intelligence CAI 2024 proceedings. }
}


 \author{\IEEEauthorblockN{Dylan Seychell}
 \IEEEauthorblockA{\textit{Dept. of AI} \\
 \textit{University of Malta}\\
 Msida, Malta \\
 dylan.seychell@ieee.org}
 \and
 \IEEEauthorblockN{Gabriel Hili}
 \IEEEauthorblockA{\textit{Dept. of AI} \\
 \textit{University of Malta}\\
 Msida, Malta \\
 gabriel.hili@um.edu.mt}
 \and
 \IEEEauthorblockN{Jonathan Attard}
 \IEEEauthorblockA{\textit{Dept. of AI} \\
 \textit{University of Malta}\\
 Msida, Malta \\
 jonathan.attard.20@um.edu.mt}
 \and
 \IEEEauthorblockN{Konstantinos Makantatis}
 \IEEEauthorblockA{\textit{Dept. of AI} \\
 \textit{University of Malta}\\
 Msida, Malta \\
 konstantinos.makantasis@um.edu.mt}
}

\maketitle

\begin{abstract}
In today's media landscape, the role of Artificial Intelligence (AI) in shaping societal perspectives and journalistic integrity is becoming increasingly apparent. This paper presents two case studies centred on Malta’s media market featuring technical novelty. Despite its relatively small scale, Malta offers invaluable insights applicable to both similar and broader media contexts. These two projects focus on media monitoring and present tools designed to analyse potential biases in news articles and television news segments. The first project uses Computer Vision and Natural Language Processing techniques to analyse the coherence between images in news articles and their corresponding captions, headlines, and article bodies. The second project employs computer vision techniques to track individuals' on-screen time or visual exposure in news videos, providing queryable data. These initiatives aim to contribute to society by providing both journalists and the public with the means to identify biases. Furthermore, we make these tools accessible to journalists to improve the trustworthiness of media outlets by offering robust tools for detecting and reducing bias.
\end{abstract}


\section{Introduction}
The world's social challenges, such as the influence of public perspectives, are being shaped as technical challenges.  In this context, the societal implications of Artificial Intelligence (AI) must be given the needed weight. Therefore, society needs to advance its ethical AI practices, as outlined in the recent Council of Europe guidelines on AI in journalism \cite{coeGuidelines2023}. The induction of bias in content is constantly challenging elections, public opinion, and democracy itself.  Journalism and the media are responsible for informing the public with fair information. Information consumers can use AI to detect potential bias while also providing insights to journalists on the content they produce.  As the influence of media expands globally, developing evidence-based techniques to promote fairness and accuracy is critical.

This paper presents two technical case studies on Malta's media market, which, despite being small, provide unique insights that can be extrapolated to similar contexts or broader landscapes.  These two projects explore the area of media analysis and provide tools to study potential bias in news articles and news segment videos, respectively.  In the first project, we demonstrate how a range of AI techniques can be used to measure the match between the content of an image in a news article with its caption, article headline and body. In the second project, we demonstrate how computer vision techniques can be used to monitor the camera time of individuals appearing in news videos while providing queriable results. These projects directly impact society by empowering journalists and the public with tools to discern biases and inaccuracies in news content.  Moreover, we aim to improve media outlets' credibility by providing tools to detect and reduce bias.

\section{Background and Literature Review}

This section discusses different types of media bias and its detection. Then, we present Vision Language Pre-training and Video Analysis techniques used in our case studies. 

\subsection{AI for Media Bias Mitigation}
Media bias can be classified into different types \cite{puglisi2015empirical}. Distortion bias occurs when facts are omitted, events are ignored, or numbers are presented as official without proper sourcing. Filtering bias involves cherry-picking facts when summarising an event to slant the news in a particular direction. Ideological bias promotes a specific political party or ideology, while spin bias aims to create attention-grabbing stories. Gatekeeping bias involves deciding which stories to report or suppress, while coverage bias entails unequal visibility given to certain topics. Lastly, statement bias influences reader sentiment through the language used. Contribution towards these biases can all be committed via clever usage of images in the article, namely the choice of image, position, size, visibility, description, and content.

Moreover, different automatic techniques exist to analyse these different biases \cite{hamborg2019automated}. For example, plagiarism detection methods can be used to identify content reuse across newspapers. News Sentiment Classifiers datasets \cite{Hamborg2021b} can also be applied to news article text. Facial Detection and Recognition help detect how much exposure is given to certain persons in news broadcasts or news articles. However, given the early days of VLP technology \cite{CGV-105}, it appears that no other study leveraged the ability of VLP models to provide a multimodal bias analysis for news content, such as the approaches presented in this paper.

\subsection{Vision Language Pre-training} 
Visual Language Pre-training (VLP) combines elements from CV and NLP for downstream uni-modal tasks while minimising the need for training a new model from the ground up \cite{chen2023vlp}. VLP models aim to understand images in relation to natural language and develop a universal abstract understanding of how humans tend to describe pictures using words. With advancements in natural language model pre-training, namely BERT \cite{devlin2018bert},  transformers have become the standard architecture powering all state-of-the-art VLP models. Some examples of these large-scale VLP models are ALIGN \cite{jia2021scaling} and BLIP \cite{li2022blip}. VLP is helpful in various applications, including Caption Generation, Image-Text Retrieval, Image Classification, and Natural Language Visual Reasoning (NLVR). VLP models generally comprise three modules: the Vision Encoder, the Text Encoder, and the Multimodal Fusion. The Vision Encoder aims to extract significant features from an input image, categorised into Region Features, Grid Features, and Patch Projection. The OD-based Region Features (OD-RF) \cite{lu2019vilbert} approach uses an object detection model to extract bounding-box features. A drawback of OD-RF is that the vision backbone is frozen during training, which restricts overall performance. On the other hand, CNN-based Grid Features (CNN-GFs) \cite{wang2021simvlm} enable end-to-end (e2e) processing. Lastly, Vision Transformers (ViT) \cite{dosovitskiy2020image} split images into patches and project them into embeddings. Text encoders such as Word2Vec \cite{mikolov2013efficient}, GloVe \cite{pennington2014glove} aim to create highly informative latent text representations. Due to the success of BERT \cite{devlin2018bert}, state-of-the-art text encoders generally employ transformers for text representation. In the multifusion module, both embeddings are combined to produce a joint representation of the image‐text pair, typically done using attention. There are two types of multimodal fusion: merged attention (used in VisualBERT \cite{li2019visualbert}, UNITER \cite{chen2020uniter}, ViLT \cite{kim2021vilt}, and GIT \cite{wang2022git}) and co-attention (used in LXMERT \cite{tan2019lxmert} and ViLBERT \cite{lu2019vilbert}). Merged attention concatenates features and passes them to the same transformer block, while co-attention feeds features into separate transformer blocks with cross-attention. 

\subsection{Media Bias Analysis in Videos}
In their paper, Lisena et al. \cite{lisena2021facerec} proposed a system for face recognition in videos using a combination of MTCNN \cite{zhang2016joint}, FaceNet \cite{schroff2015facenet}, and SVM classifiers. The system obtained images through crawlers, extracted face embeddings using MTCNN and FaceNet, and used an SVM classifier to identify known faces with confidence scores. Frame-level analysis was performed, with face detection, cropping, alignment, and recognition for each frame. The SORT algorithm was used for face tracking, and a weighted average algorithm was employed to label faces in frame sequences. Hierarchical clustering was used to group similar face encodings, and a ground truth dataset was created for evaluation using news videos. Although the system demonstrates good performance, upon further investigation, it was noticed that a limited number of individuals were used within the face recognition system, which is important to consider since the error rate might increase with the number of individuals. Such systems can also be further optimised using Saliency Ranking techniques \cite{seychellSARA} that can guide the attention towards more salient regions of the scene.

Optical Character Recognition (OCR) is a process used to extract text from images, involving various steps such as preprocessing, segmentation, feature extraction, classification, and post-processing techniques such as NLP and dictionary-based approaches  \cite{islam2017survey}. Challenges in OCR include dealing with different fonts or handwriting, text-background contrast, and indistinguishable characters, which can be addressed through consistent font usage and appropriate preprocessing methods \cite{mithe2013optical}. Several studies have utilised OCR in analysing news videos, with Tesseract \cite{tesseract-ocr} being a powerful open-source OCR tool used in multiple studies \cite{ wattanarachothai2015key}. Wattanarachothai et al. \cite{wattanarachothai2015key} presented a novel approach for video content retrieval using text-based techniques involving key frame extraction, text localization, and keyword matching. Their method utilised Maximally Stable Extremal Region (MSER) features for key frame extraction, clustering MSERs for text localisation, and Tesseract OCR engine for text recognition. Experimental results showed the effectiveness of the MSER feature in video segmentation and highlighted improved precision and recall compared to alternative methods.

\section{Proposed AI in Journalism Tools}
  
This section presents two novel AI systems designed to help consumers better understand news content while serving as a tool for journalists.

\subsection{News Articles Analysis} \label{sec:Met-articles}
The news article analysis system uses data from six online newspapers: The Times of Malta (ToM), The Shift (TS), Malta Today (MT), The Malta Independent (MI), Malta Daily (MD), and Gozo News (GN). 

\begin{figure}[h]
    \centering
    \includegraphics[scale=0.72,keepaspectratio]{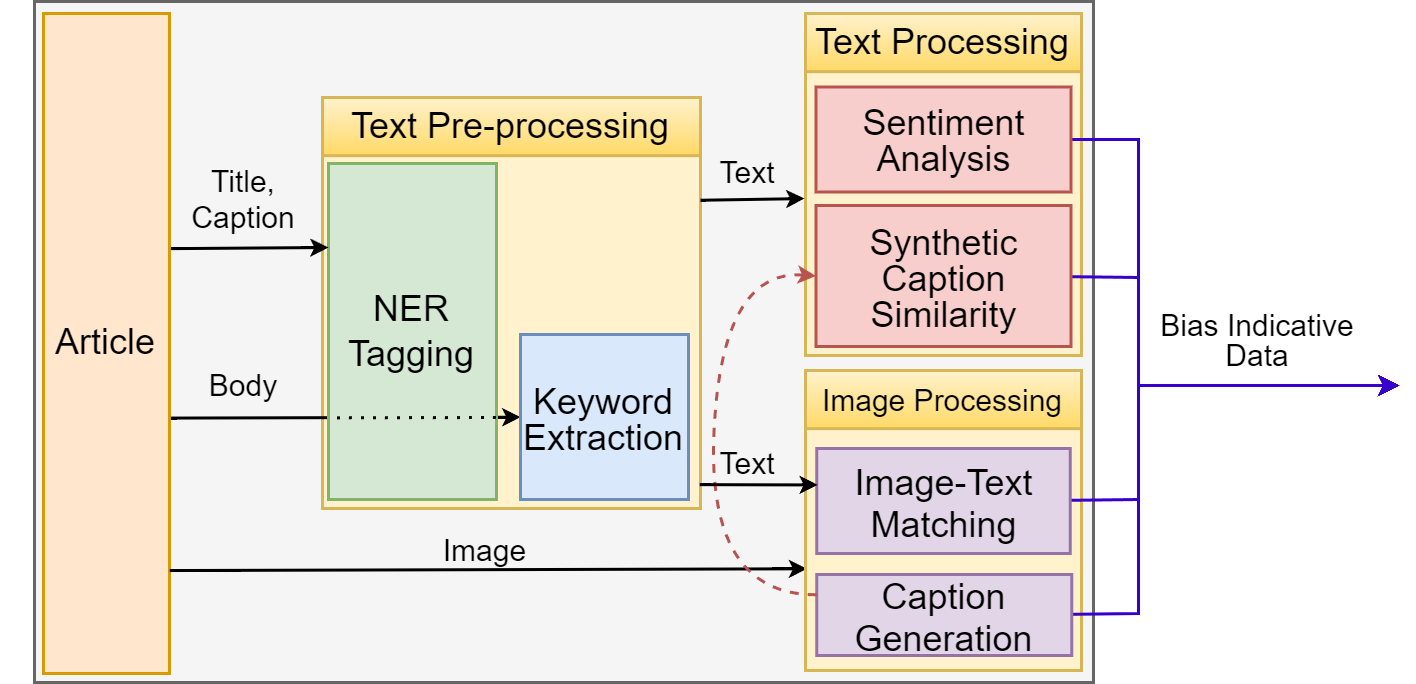}
    \caption{Diagram showing the data transformation pipeline of the news articles data.}
    \label{fig:methodology_pipeline}
\end{figure}

The dataset includes 800 news articles from each online newspaper from the 5th of April, 2023. All articles from ToM included at least one caption describing an image. 1.75\% of MT articles (14) did not have at least one caption, while this percentage was 51.00\% (408) for TS. The articles from the other newspapers did not include a caption with the accompanying images. The title, article body, and image-caption pairs of every article are extracted to local storage, where six main transformations will be performed on the data: Named Entity Recognition (NER) Tagging, Keyword Extraction, Sentiment Analysis, Caption Generation, Synthetic Caption Similarity, and Image-Text Matching (ITM). Caption Generation and ITM are performed using BLIP with a ViT-L/16 \cite{wu2020visual} vision backbone. These six transformations aim to extract bias-indicative insights from the news article data, which can be analysed and visualised by media bias researchers. Figure \ref{fig:methodology_pipeline} presents the high-level architecture for this system.

First, NER tagging was performed on the raw text. A list of Maltese town names was acquired from Wikipedia to improve the NER tagger's performance. Any mention of the town names in the articles was replaced with \say{Tokyo}, as the NER tagger sometimes would not correctly label them as a location, perhaps due to their low frequency, if any, in the NER tagger's training data. Then, keyword extraction was performed on the body of the NER-tagged article. In total, 20 keywords and five single-or-double-worded keywords were extracted from the body. The Target-dependent Sentiment Classifier (TSC) model utilised \cite{Hamborg2021b} calculated the sentiment score for any mention of the NER tags \say{Person}, \say{Organisation}, and \say{Location} by looking at the previous and next words in the sentence. The positive, neutral, and negative scores are each averaged per article to produce an aggregate value for the current field (title, caption, or body). If none of the NER tags are present in the text, TSC is performed on the entire text with zero leading or trailing tokens. The images of the articles are passed to the VLP model BLIP \cite{li2022blip} for caption generation. First, the images are resized to 640$\times$640px and normalised using the default mean values of the original study. The captions are chosen using nucleus sampling \cite{holtzman2019curious} with the sum probability of chosen tokens equal to 0.9. The length of the generated caption is between 5 and 20 words. The Python package \textit{sentence-transformer 2.2.2} \cite{reimers-2019-sentence-bert} makes use of the \textit{all-MiniLM-L6-v2} pre-trained model to calculate a similarity score between the generated caption and the NER-tagged title, caption, and body keywords. Lastly, Image-Text Matching (ITM) is performed using BLIP. The NER-tagged title, caption, and body keywords are matched to the accompanying images of the article and two similarity scores (softmax output and cosine similarity) are extracted per match.

\subsection{News Video Analysis}

This section presents an automated system for monitoring news videos using computer vision techniques. The technical contributions are as follows. Firstly, we curated and annotated a new dataset of 215 news segment videos, comprising 8.48 hours, from TVM news containing named persons in captions. Third-party annotators carefully annotated the dataset, providing accurate timestamps, names, and flags indicating name visibility and presenter status. These annotations allow for a detailed analysis of the system's results and performance regarding OCR and person detection. The dataset was used to train and evaluate the system. The resolution of the videos ranged up to 1280$\times$720px.

A pipeline for automatically extracting depicted persons' names using optical character recognition on caption frames was designed and built.  This system was also built with privacy by design.  For this reason, the system only extracts information that appears on the news video without access to external systems or sources.  Moreover, it is built to be tuned to only store information about individuals holding public office to minimise any harmful application.  Nonetheless, the capability for automated naming is novel. Building on this, we present our machine-learning approach for detecting, tracking and temporally localising individuals throughout the videos. This facilitates the automated generation of camera-time statistics.

\begin{figure}[!htb]
    \centering  \includegraphics[scale=0.11,keepaspectratio]{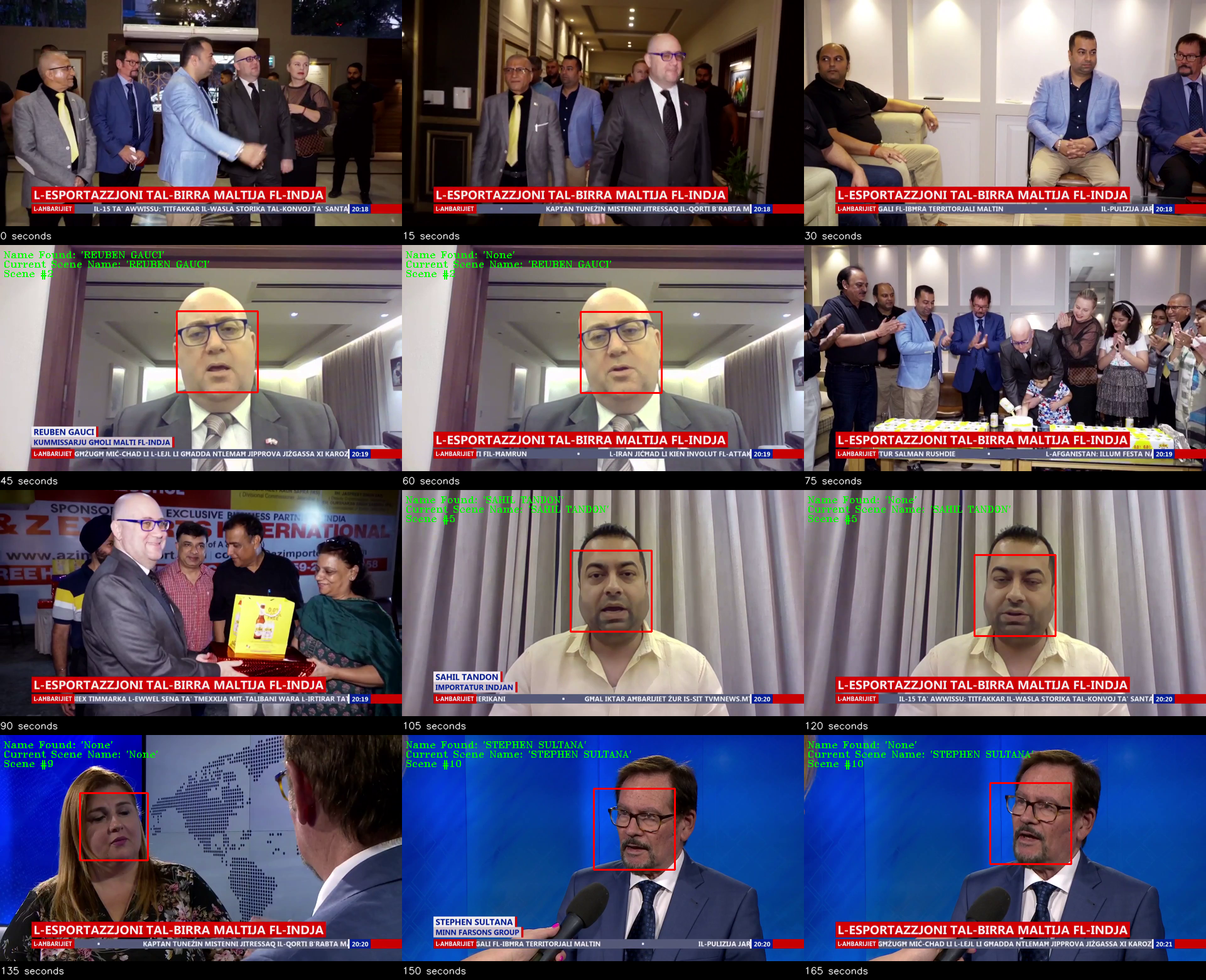}
    \caption[Frame timeline example]{A frame timeline showing three people who appeared, along with the real‐time
analysis annotations. The generated analysis timeline can be seen in the leftmost timeline in Figure \ref{fig:analysis_timeline}. \label{fig:frames}}
\end{figure}

The Video Analysis system comprises sub-components: face detection, recognition, object tracking, and name extraction. These work together to extract faces, track individuals, and retrieve names to create a timeline.  Figure \ref{fig:frames} shows a selection of frames from the auxiliary output provided by the system, which is intended to provide transparency and accountability when supporting the timeline illustrated in Figure \ref{fig:analysis_timeline}.

\subsubsection{Name Extraction\label{meth:name_extraction}}

The name extraction component is tailored to extract names from videos with a specific structure. The system utilises pre-processing operations, including image closing, opening, and HSV colour thresholding. Box-like contours are identified through conditional statements and extensive testing. The preprocessed image is then processed using Pytesseract\footnote{The Tesseract Library is available from https://pypi.org/project/pytesseract/} OCR, which is optimised to recognise structured text, including the white box with blue text found in the videos. The most common non-empty text in a scene is selected as the individual's name to handle text transition animations. The original name is used if a match is found in the database; otherwise, the extracted name is saved in a local database. This approach successfully extracted names from most name captions in the videos.

\subsubsection{Face Detection, Encoding, and Recognition\label{meth:face_det_enc_recog}}
The distance between the face and the database encodings is calculated for each frame to match a detected face with a known individual. The average distance across all frames is compared to a given threshold in an adaptation of FaceRec \cite{lisena2021facerec}. However, an issue arises when adding an individual's facial encoding to the database due to multiple face encodings extracted from multiple frames. Different approaches were employed to select the most suitable facial encoding, including taking the first and middle frames.  This was based on the work inspired by Gao et al.'s work on scene detection \cite{gao2002unsupervised} while calculating the average of the extracted encodings. These strategies contributed to improving facial recognition accuracy.

\subsubsection{Face Tracking\label{meth:face_tracking}}

\begin{figure}[!htb]
    \centering
   \includegraphics[scale=0.29,keepaspectratio]{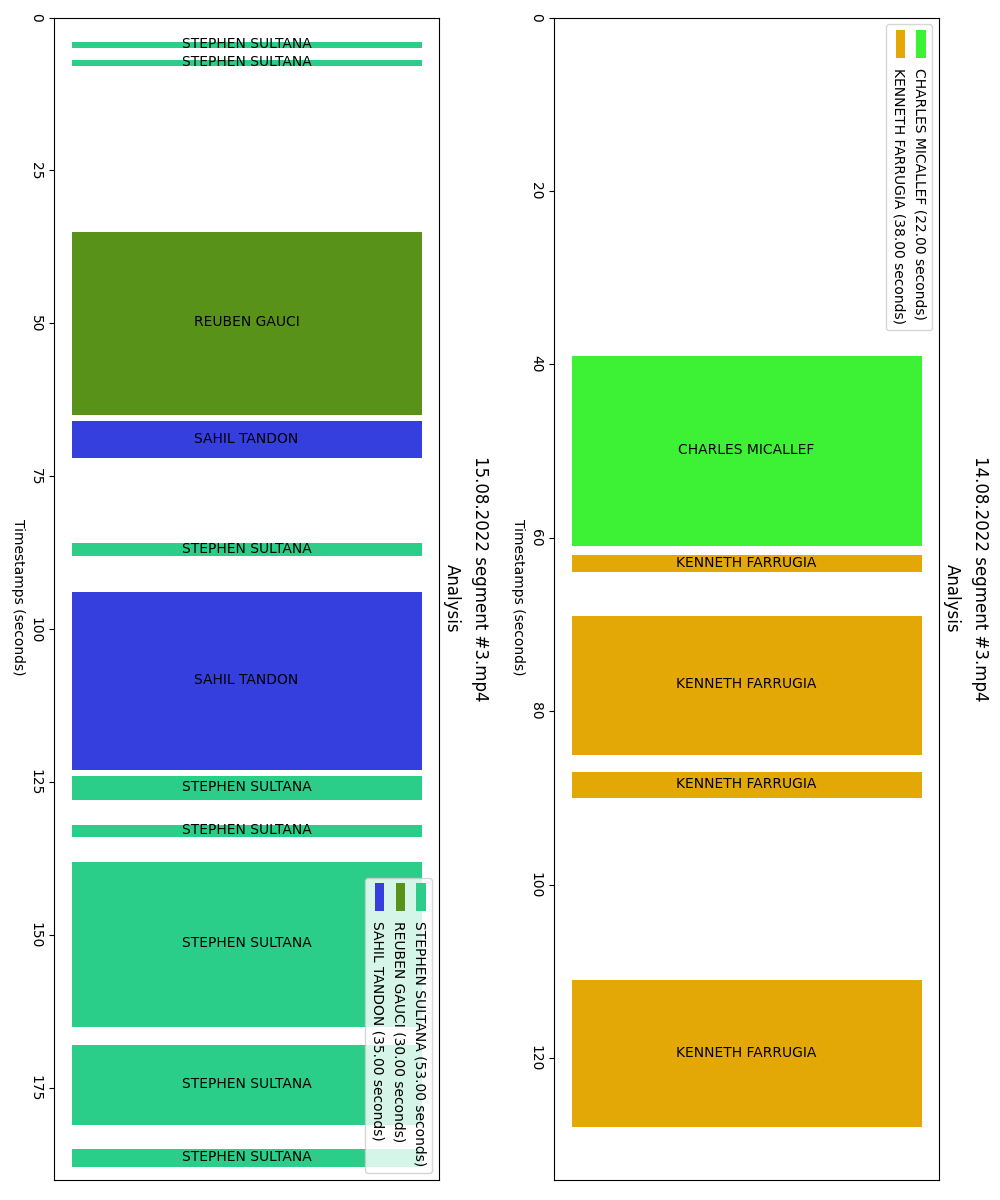}
   \caption[Analysis timeline example]{The analysis timeline showing occuring individuals by their names inclduing their timestamps and durations.\label{fig:analysis_timeline}}
\end{figure}

Initially, attempts at scene detection were employed to segment videos into shots, but they were found to be unreliable due to sudden movements and prolonged shots that featured multiple people or individuals for only a brief moment. As a result, a tracking approach \cite{lisena2021facerec} was adopted to improve accuracy. The Kernelised Correlation Filters (KCF) \cite{henriques2014high} algorithm was used for scene segmentation. The KCF algorithm initialises a bounding box around a detected face and tracks it across subsequent frames. The tracker automatically adapts to changes in scale and position, ensuring the bounding box stays aligned with the tracked face. The tracker was exclusively used to indicate the start and end scenes of the individual used to generate the timeline visualisation presented in Figure \ref{fig:analysis_timeline}.

\section{Evaluation}

\subsection{Evaluating the News Article Analysis}
This section presents the evaluation of the news article analysis system.  The machine-learning data-transformation pipeline was evaluated on a total of 4,800 articles from six Maltese newspapers listed in Section \ref{sec:Met-articles}. One of the goals of the evaluation was to examine whether newspapers artificially add to an article's perceived sentiment by intentionally using irrelevant images. The correlation coefficients in Table \ref{tab:itmXkw_all_sentiment} show whether opinionated newspapers tend to have poorer image relevancy to the article body or vice-versa. A positive value is related to increased image relevancy as the article has more positive/neutral/negative scores. A negative value means image relevancy decreases. From the results obtained, no correlation was found between these two variables.

 \begin{table}[h]
            \centering
            \caption{Correlation coefficients between sentiment and image similarity for different newspapers' articles' body.}
            \begin{tabular}{c||c|c|c}
                 \textbf{Newspaper}&\textbf{Positive}&\textbf{Neutral}&\textbf{Negative}  \\\hline
                 ToM   &0.12&0.02&-0.12 \\ 
                 TS        &0.04&0.02&-0.05 \\ 
                 MT      &0.15&-0.05&-0.09 \\ 
                 MI&-0.09&-0.02&-0.09 \\ 
                 MD      &0.08&-0.02&-0.1 \\ 
                 GN        &0.09&-0.07&-0.04 \\ 
            \end{tabular}
            
            \label{tab:itmXkw_all_sentiment}
        \end{table}

Another goal of the evaluation was to investigate the relationship between the images and article text.  Table \ref{tab:final_table} summarises the main findings resulting from this experiment. The newspaper with the highest similarity score to the synthetic captions (SSC) was the Times of Malta (ToM). Malta Daily (MD) performed the best regarding Image-Text Matching (SI). It was also noted that Gozo News (GN) articles have the least negative sentiment overall, while The Shift (TS) ranked one of the lowest in almost all metrics.

    \begin{table}[h]
        \centering
        \caption{A table of the aggregated mean results of the demonstration. All values represent a percentage.}
        \begin{tabular}{l||p{0.7cm}|p{0.7cm}|p{1cm}|p{1cm}|p{1cm}}
            \textbf{Newspaper} & \textbf{SSC}   & \textbf{SI}    & \textbf{Positive Sent.} & \textbf{Neutral Sent.} & \textbf{Negative Sent.}\\\hline
            ToM                & \textbf{15.99} & 30.99          &  13.87                  & 65.55                  & 20.58                  \\
            TS                 & \textbf{9.27}  & 20.78          &  11.71                  & 62.68                  & 25.60                  \\
            MT                 & 10.80          & 24.15          &  16.79                  & 67.03                  & 16.18                  \\
            MI                 & 10.96          & 29.92          &  21.27                  & 62.56                  & 16.17                  \\
            MD                 & 13.46          & \textbf{50.79} &  \textbf{28.81}         & 58.72                  & 12.47                  \\
            GN                 & 13.58          & 38.76          &  25.17                  & \textbf{69.46}         & \textbf{5.37}          \\
        \end{tabular}
        \label{tab:final_table}
    \end{table}

\begin{table*}[hbt!]
    \centering
    \caption[Video Analysis Results]{The Video Analysis system's performance is evaluated using precision (P), recall (R), F1-score (F1), and accuracy (A) metrics for name extraction and face recognition. The MAE measures duration accuracy together with the processing time.}
    \label{tab:results}
    \scalebox{1}{
        \begin{tabular}{|l||c|c|c|c||c|c|c|c||c||c|}
            \hline
            
            \multicolumn {1}{|c||}{\textbf{Variation}} &
            \multicolumn {4}{c||}{\textbf{Name extraction}} & 
            \multicolumn {4}{c||}{\textbf{Face recognition}} & 
            \multicolumn {1}{c||}{\textbf{MAE}} &
            \multicolumn {1}{c|}{\textbf{Time}}
            \\ \hline
            
            \multicolumn {1}{|c||}{\textbf{metrics}} & \textbf{P} & \textbf{R} & \textbf{F1} & \textbf{A} & \textbf{P} & \textbf{R} &\textbf{ F1} & \textbf{A} & \textbf{(secs)} & \textbf{(hrs)}
            \\ \hline
    
            \multicolumn {11}{|c|}{\textbf{Scene Detection}}
            \\ \hline
            
            Skips (1, 2) &
            0.78  & 0.67  & 0.72  & 0.56  & 
            0.81  & 0.65  & 0.72  & 0.57  & 
            10.20 &
            4.71
            \\ \hline
    
            Skips (2, 1) &
            0.81  & 0.67  & 0.73  & 0.58  & 
            0.84  & 0.65  & 0.73  & 0.57  & 
            05.73 &
            4.27        
            \\ \hline
    
            Skips (2, 2) &
            0.79  & 0.66  & 0.72  & 0.56  & 
            0.82  & 0.65  & 0.72  & 0.56  & 
            09.49 &
            2.95
            \\ \hline
    
            Skips (2, 3) &
            0.81 & 0.67  & 0.73  & 0.58  & 
            0.82  & 0.63  & 0.71  & 0.56  & 
            08.45 &
            2.42
            \\ \hline
    
            Skips (3, 2) &
            0.81  & 0.62  & 0.70  & 0.54 & 
            0.82  & 0.60  & 0.70  & 0.53  & 
            05.93 &
            1.94
            \\ \hline
    
            Def. (2, 2) &
            \underline{\textbf{0.83}}  & \underline{\textbf{0.72}} & \underline{\textbf{0.77}}  & 0.62  & 
            0.84  & 0.69  & 0.76  & 0.61  & 
            09.87 &
            4.51
            \\ \hline
    
            \multicolumn {11}{|c|}{\textbf{Face Tracking}}
            \\ \hline
    
    
    
            Fir. (2) &
            0.81  & 0.66  & 0.73  & 0.57  & 
            0.83  & 0.65  & 0.73  & 0.57  & 
            07.40 &
            2.46
            \\ \hline
    
            Mid. (0.5) &
            0.78  & \underline{\textbf{0.72}}  & 0.75  & 0.60  & 
            0.82  & 0.68  & 0.74  & 0.59  & 
            \underline{\textbf{04.22}} &
            7.59
            \\ \hline
    
            Mid. (2) &
            0.81 & 0.68  & 0.74  & 0.59  & 
            0.83  & 0.63  & 0.72  & 0.56  & 
            06.89 &
            1.98
            \\ \hline
    
            Avg. (0.5) &
            0.78  & 0.70  & 0.74  & 0.58  & 
            0.80  & 0.66  & 0.72  & 0.57  & 
            04.74 &
            7.20
            \\ \hline
    
            Avg. (1) &
            0.80  & 0.70  & 0.75  & 0.60  & 
            0.83  & 0.68  & 0.75  & 0.60  & 
            05.72 &
            4.48
            \\ \hline
    
            Avg. (2) &
            0.81  & 0.66  & 0.72  & 0.57  & 
            0.83  & 0.61  & 0.70  & 0.54  & 
            07.18 &
            \underline{\textbf{1.76}}
            \\ \hline
    
            Def. (1) &
            \underline{\textbf{0.83}}  & \underline{\textbf{0.72}} & \underline{\textbf{0.77}} & \underline{\textbf{0.83}}  & 
            \underline{\textbf{0.86}}  & \underline{\textbf{0.71}}  & \underline{\textbf{0.78}} & \underline{\textbf{0.63} } & 
            05.47 &
            5.91
            \\ \hline
            
            Def. (2) &
            0.82  & 0.70 & 0.75 & 0.60  & 
            0.85  & 0.67  & 0.75  & 0.60 & 
            07.03 &
            4.79
            \\ \hline
        
        \end{tabular}
    }
\end{table*}

\subsection{Evaluating the News Video Analysis}

This section evaluates the system for automatic analysis of news videos, focusing on its strengths, weaknesses, and performance. The Video Analysis component, which forms the system's core, is essential to evaluate. While previous studies have explored news videos \cite{gao2002unsupervised, lee2017strategy, qi2000integrating, sato1998video, yang2004naming} and face recognition in videos \cite{lisena2021facerec, wang2009video}, few have addressed live face recognition and name extraction in news videos simultaneously. Therefore, each sub-component will be separately assessed to determine the system's overall performance.

A comprehensive selection of metrics was employed to assess each sub-component, including name extraction, face recognition, facial tracking, and analysis duration, to evaluate the system's performance.  Testing was conducted to identify optimal parameters and methods, and the results are presented in Table \ref{tab:results}, which provides an overview of the achieved performance with different variations. An incremental database test was also conducted to assess the system's scalability. Evaluating these metrics offers insights into the system's performance and areas for improvement.

\subsubsection{Metrics\label{eval:metrics}}

The performance of the name extraction sub-component is evaluated using a confusion matrix, comparing the system's identified names against the names in the validation set per video. Using the confusion matrices, metrics such as precision, recall, F1 score, and accuracy score are calculated to assess the name extraction and face recognition performances. To evaluate the accuracy of the scene prediction sub-component or tracker for each person, the Mean Absolute Error (MAE) is calculated. The MAE duration represents the average difference in seconds between the predicted and actual durations for each individual from the annotated dataset. A smaller MAE duration indicates better predictions.

\subsubsection{Results\label{eval:results}}

Table \ref{tab:results} presents the performance of the Video Analysis system for the tested variations. The table is divided into two sections: one for scene detection methods and the other for the latest method using face tracking. The numbers in parentheses indicate the seconds skipped between frames during analysis. For scene detection methods, two values are displayed: the seconds skipped during analysis and the seconds skipped during scene detection. `Skips' indicates the use of the same default parameters. 'Def.' signifies the default resolution, while 640$\times$360px was used to speed up frame analysis in other cases. `Fir.', `Mid.', and `Avg.' correspond to the first, middle, and average face encodings, respectively, as explained in Section \ref{meth:face_det_enc_recog}. In scene detection, only the first encountered face encoding was used for each individual. The metrics P (Precision), R (Recall), F1 (F1-score), and A (Accuracy), MAE (Mean Absolute Error) were calculated as described in Section \ref{eval:metrics}. The Time column represents the hours required to analyse all the 'News Video Segments' videos. Multiprocessing was employed in some analyses to reduce processing time. The best metric achieved for each calculation is highlighted in bold and underlined.

During the investigation, the name extraction process encountered several issues: detecting names with punctuation marks, misidentifying the right side of the caption box as `L,' misinterpreting `L' as the end of the caption box, exceeding frame view limitations, and skipping frames with names during video analysis. Addressing these issues is crucial to enhancing extraction effectiveness. The results show that a one-second interval for video analysis achieved the best scene detection and face tracking performance, consistent with \cite{lisena2021facerec}. Lower intervals in scene detection improved accuracy, while larger intervals in face tracking led to higher errors. When compared to Scene Detection and Face Tracking, Face Tracking achieved a lower MAE value, indicating better results in selecting individual timestamps. The performance of face recognition does not vary significantly with different face encoding selections. As suggested in \cite{gao2002unsupervised}, the middle face encoding generally yields slightly better results. However, relying solely on the middle or first face encoding might return lower accuracy if the face is unclear or poorly represented. The results in Table \ref{tab:results} align with existing work \cite{gao2002unsupervised} \cite{lisena2021facerec} where the optimal configuration for the Video Analysis system involves using face tracking and analysing frames at 1-second intervals. The default resolutions performed slightly better, but the longer processing time does not justify this marginal advantage. Using a resolution of 640$\times$360px is recommended for faster analysis. Overall, the results demonstrate promising performance with potential areas for improvement in each sub-component.

\section{Conclusion}

This paper presented two technical case studies of applied AI in journalism and media, each technique presenting its novelty. We believe transparently surfacing potential distortions in news content allows media professionals of good faith opportunities to focus their efforts accordingly towards fairness, in line with the Council of Europe’s recommendations on disclosure of AI use \cite{coeGuidelines2023}.  We plan for future work to expand the reach of these projects and validate them in broader contexts.

\bibliographystyle{IEEEtran}
\bibliography{refs}

\end{document}